\documentclass[aps,prl,twocolumn,showpacs,amsmath,amssymb,groupedaddress]{revtex4-1}
\usepackage{graphicx}
\begin{document}
\title{Spin-Orbit Coupled Bose Gases at Finite Temperatures}
\author{Renyuan Liao$^1$, Oleksandr Fialko$^2$}
\affiliation{$^1$College of Physics and Energy, Fujian Normal University, Fuzhou 350108, China}
\affiliation{$^2$Centre for Theoretical Chemistry and Physics, Massey University, Auckland 0632, New Zealand}
\date{\today}
\begin{abstract}
Spin-orbit coupling is predicted to have dramatic effects on thermal properties of a two-component atomic Bose gas.
We show that in three spatial dimensions it lowers the critical temperature of condensation and enhances thermal depletion of the condensate fraction.
In two dimensions we show that spin-orbit coupling destroys superfluidity at any finite temperature, modifying dramatically the cerebrated
Berezinskii-Kosterlitz-Thouless scenario. We explain this by the increase of the number of low energy states induced by spin-orbit coupling, enhancing the role
of quantum fluctuations.
\end{abstract}
\maketitle
There are numerous phenomena in a wide range of quantum systems, ranging from condensed matter to atomic and nuclear physics, where spin-orbit coupling (SOC) plays
an important role. Recently discovered new class of topological insulators, quantum spin Hall effect~\cite{XIA} and Majorana fermions~\cite{MOU12} rely on SOC and are
expected to retain their quantum properties up to room temperature. However, the electronic systems can not be easily controlled and the details of SOC are usually not known.
Therefore, it is a difficult task to manipulate such systems. In contrast, ultracold atoms have been demonstrated to be a remarkable platform for emulation of various
condensed matter phenomena due to their ability to be easily manipulated at will~\cite{ZOL05}. The pioneering experimental realization of synthetic gauge fields and
SOC~\cite{LIN,BLO11,CHE12,ZHA12,ZWI12} is defining a new dimension in exploring quantum many-body systems with ultracold atomic gases. The engineered SOC (with equal
Rashba and Dresselhaus strength) in a neutral atomic Bose-Einstein condensate was achieved by  dressing two atomic spin states with a pair of lasers.  It allows to
study the rich physics of SOC effects in  bosonic systems~\cite{STA08,ZHA10,CHA11,TIN11,YUN}, which have not been explored before. Recently, methods to generate pure
Rashba type SOC have been suggested~\cite{AND}. Its realization will make it possible to study rich ground state physics proposed in fermionic~\cite{FER} and bosonic
systems~\cite{ZHA10,BOSE}, of which many properties have no condensed matter analogues.

SOC leads to a huge degeneracy of the ground state of a single particle~\cite{LIU12}. This enhances the role of quantum fluctuations making condensation of non-interacting
bosons not possible~\cite{LIU12,QI13}. However, it has been shown that interactions among atoms stabilize condensation~\cite{BAR12,CUI12,BAY12}. The role of quantum
fluctuations is especially essential in two dimensions destroying condensation but not necessary superfluidity. This yields, in particular, the celebrated
Berezinskii-Kosterlitz-Thouless (BKT) phase transition in two dimensions with a critical temperature separating superfluid and normal phases. How do the quantum
fluctuations in the presence of SOC affect the BKT phenomenon? What is the effect of SOC on thermal properties of a Bose condensate? These questions shall be addressed
in this Letter. Previous theoretical studies have been focused mainly on the ground state properties of interacting SOC quantum gases, leaving the experimentally
relevant physics at finite temperatures intact. In the light of the recent experiment~\cite{PAN13} on a finite-temperature phase diagram of SOC Bose gases,
our present study is an interesting and urgent task.

According to the Mermin-Wagner theorem long-range order at finite temperature does not exist in one spatial dimension.
In this Letter, we shall present studies of a SOC two-component atomic Bose gas at finite temperature in two and three spatial dimensions.
The interplay between quantum and thermal fluctuations in the presence of SOC is shown to yield dramatic modifications of the familiar
physics. First, by resorting to the Popov approximation, we develop a formalism suitable for treating the system at finite temperature in three dimensions.
Within this formalism, we find that the SOC greatly suppresses the critical temperature of condensation and enhances thermal depletion of the condensate. We then derive an effective theory suitable to study the celebrated BKT phase transition in two dimensions.
The BKT transition temperature, in contrast to the previous case, is shown to drop to zero in the presence of SOC.

We consider a three-dimensional homogeneous two-component Bose gas with an isotropic in-plane (x-y plane) Rashba spin-orbit coupling, described by the following grand canonical Hamiltonian in real space:
\begin{eqnarray}
   H&=&\int d\mathbf{r}\left[\sum_\sigma\psi_\sigma^\dagger\left(-\frac{\hbar^2\nabla^2}{2m}-\mu\right)\psi_\sigma+(\psi_\uparrow^\dagger \hat{\cal{R}}\psi_\downarrow+{\rm h.c.})\right.\nonumber\\
   &+&\left.\sum_\sigma\frac{g_{\sigma\sigma}}{2}(\psi_\sigma^\dagger\psi_\sigma)^2+ g_{\uparrow\downarrow}\psi_\uparrow^\dagger\psi_\uparrow\psi_\downarrow^\dagger\psi_\downarrow\right].
 \end{eqnarray}
Here, $\psi_{\sigma}$ is a Bose field satisfying the usual commutation relation $[\psi_{\sigma}({\bf r}), \psi^\dagger_{\sigma'}({\bf r}')]=i\hbar\delta_{\sigma\sigma'} \delta({\bf r}-{\bf r}')$,
the spin index $\sigma=\uparrow,\downarrow$ denotes two pseudo-spin states of the Bose gas with atomic mass $m$. The inter-particle interaction $g_{\sigma\sigma^\prime}$ is related to the two-body scattering length $a_{\sigma\sigma^\prime}$ as
$g_{\sigma\sigma^\prime}=4\pi\hbar^2a_{\sigma\sigma^\prime}/m$. For simplicity, we shall assume that the interactions between like-spin particles are the same,
$g_{\uparrow\uparrow}=g_{\downarrow\downarrow}=g$. The chemical potential $\mu$ is introduced to fix the total particle number density. The Rashba spin-orbit coupling
is described by the operator $\hat{\cal{R}}=\lambda (\hat{p}_x-i\hat{p}_y)$ with $\lambda$ being the coupling strength.
Throughout the rest of this paper, we set $\hbar=2m=k_B=1$ and define $n^{1/3}$ as a momentum scale, while $n^{2/3}$ as an energy scale.
For the system to be weakly interacting, we set $g=0.1n^{-1/3}$.

The Hamiltonian of a non-interacting system is diagonalized in the helicity basis after the Fourier transform of the fields
$\psi_{\sigma}({\bf r})=1/\sqrt{L^3}\sum_{\bf q}\psi_{\sigma}({\bf q})\exp(-i{\bf q r})$. The gas is assumed to be in a box with size $L$.
This results in two branches of spectrum
$E_\mathbf{q}^{\pm}=\mathbf{q}^2-\mu\pm \lambda q_\perp$, where $q_\perp$ is the magnitude of the in-plane momentum. The lowest energy state is therefore
infinitely degenerate, sitting on the circle $q_\perp=\lambda/2$ in the plane $q_z=0$. This increases the low energy density of states with dramatic
implications on the thermal properties of the Bose gas to be explored below. For an interacting system, earlier mean-field study~\cite{ZHA10} found that there exist
the plane wave phase (PW) for $g\ge g_{\uparrow\downarrow}$ and the striped phase for $g<g_{\uparrow\downarrow}$. The PW phase is the result of condensation at a single
finite momentum state breaking explicitly the rotational symmetry, while the striped phase represents a coherent superposition of two condensates at two opposite momenta.

Within the framework of the functional field integral, the partition function of the system
is~\cite{SIM06} $\mathcal{Z}=\int {\cal{D}}[\psi^*,\psi]\exp(-S[\psi^*,\psi])$ with the action
$S=\int_0^\beta d\tau\left[\sum_\sigma\psi_\sigma^*\partial_\tau\psi_\sigma+H(\psi^*,\psi)\right]$,
where $\beta=1/T$ is the inverse temperature. Here for simplicity we restrict ourself to study the PW phase, as analogous treatment of the striped phase
is more involved and will be addressed elsewhere. Our choice of the PW phase is also motivated by the fact that the striped phase has not been realized
experimentally yet. We further assume that the condensation occurs at momentum $\mathbf{\kappa}=(\lambda/2,0,0)$. Without loss of generality, the condensate
wavefunction can be chosen as $(\phi_{0\uparrow},\phi_{0\downarrow})=\sqrt{n_0}(1,-1) e^{i\lambda x/2}$,
with $n_0$ being the condensate density for either species. We split the Bose field into the mean-field part $\phi_{0\sigma}$ and the fluctuating part
$\phi_{\mathbf{q}\sigma}$ as $\psi_{\mathbf{q}\sigma}=\phi_{0\sigma}\delta_{\mathbf{q}\kappa}+\phi_{\mathbf{q}\sigma}$.
After substitution, the action can be formally written as $S=S_0+S_{f}$, where $S_0=\beta L^3\left[-2(\kappa^2+\mu)n_0+(g+g_{\uparrow\downarrow})n_0^2\right]$
is the mean-field contribution and $S_{f}$ denotes a contribution from fluctuating fields.
At this point, the action is exact. However, it contains terms of cubic and quartic orders in fluctuating fields.
To deal with such an action, one must resort to some sort of approximation. The celebrated Bogoliubov approximation for BEC is valid strictly at zero temperature.
At finite temperatures, the self-consistent Hartree-Fock-Bogoliubov (HFB) approximation gives a gapped spectrum~\cite{GRI96},
violating the Hugenholtz-Pines theorem~\cite{HUG59} and the Goldstone theorem~\cite{GOL62},
which results from the spontaneous symmetry breaking of U(1) gauge symmetry.
We choose the Popov theory~\cite{POP01} yielding a gapless spectrum and therefore it is more suitable for treating finite-temperature Bose gases.


Under the Popov approximation, which takes into account interactions between excitations~\cite{AND04}, the terms with three and four fluctuating fields in the action
are approximated as follows (neglecting anomalous average): $\phi_\sigma^*\phi_\sigma\phi_\sigma\approx2\langle\phi_\sigma^*\phi_\sigma\rangle\phi_\sigma$, $(\phi_\sigma^*\phi_\sigma)^2\approx4\langle\phi_\sigma^*\phi_\sigma\rangle\phi_\sigma^*\phi_\sigma$
and $|\phi_\uparrow\phi_\downarrow|^2\approx\langle\phi_\uparrow^*\phi_\uparrow\rangle\phi_\downarrow^*\phi_\downarrow+
\langle\phi_\downarrow^*\phi_\downarrow\rangle\phi_\uparrow^*\phi_\uparrow$. Within the Popov approximation
we also require the first order term to vanish, which fixes the chemical potential as $\mu=\mu_p=-\kappa^2+(g+g_{\uparrow\downarrow})n_0+(2g+g_{\uparrow\downarrow})n_f$,
where $n_f \equiv \langle\phi_\sigma^*\phi_\sigma\rangle$ is the density of particles of either component excited out of the condensate.
We define a four-dimensional column vector $\Phi_{\mathbf{q}n}=(\phi_{\kappa+\mathbf{q},n\uparrow},\phi_{\kappa+\mathbf{q},n\downarrow},
\phi^{\ast}_{\kappa-\mathbf{q},n\uparrow},\phi^{\ast}_{\kappa-\mathbf{q},n\downarrow})$, whose components are defined through the Fourier transform
$\phi_{\sigma}({\bf r},\tau)=1/\sqrt{L^3\beta}\sum_{{\rm q},n}\phi_{{\bf q},n\sigma}{\rm exp}[i(\mathbf{q}\cdot \mathbf{r}-w_n \tau)]$, where $w_n=2n\pi/\beta$ is the bosonic Matsubara frequencies.
Retaining terms of zeroth and quadratic orders in the fluctuating fields we can then bring the fluctuating part of the action into the compact form
$S_f\approx\sum_{{\bf q},n}\frac{1}{2}\Phi_{\mathbf{q}n}^*\mathcal{G}^{-1}(\mathbf{q},iw_n)\Phi^T_{\mathbf{q}n}-\beta\sum_\mathbf{q}\epsilon_{-\mathbf{q}}$, where
$\epsilon_\mathbf{q}=({\bf q}+\kappa)^2-\mu+(2g+g_{\uparrow\downarrow})(n_0+n_f)$ and the inverse Green's function $\mathcal{G}^{-1}(\mathbf{q},iw_n)$ reads
\begin{widetext}
\begin{eqnarray}
    \mathcal{G}^{-1}(\mathbf{q},iw_n)=
    \begin{pmatrix}
       -iw_n+\epsilon_\mathbf{q}   &   {\cal R}_\mathbf{\kappa+q}-g_{\uparrow\downarrow}n_0 & gn_0 & -g_{\uparrow\downarrow}n_0\\  {\cal R}_\mathbf{\kappa+q}^*-
g_{\uparrow\downarrow}n_0& -iw_n+\epsilon_\mathbf{q}& -g_{\uparrow\downarrow}n_0& gn_0\\ gn_0 & -g_{\uparrow\downarrow}n_0
& iw_n+\epsilon_{-\mathbf{q}} &  {\cal R}_\mathbf{\kappa-q}^*-g_{\uparrow\downarrow}n_0\\ -g_{\uparrow\downarrow}n_0& gn_0 & {\cal R}_\mathbf{\kappa-q}-g_{\uparrow\downarrow}n_0
& iw_n+\epsilon_{-\mathbf{q}}
    \end{pmatrix},
\end{eqnarray}
\end{widetext}\label{eq:Green}
where ${\cal R}_\mathbf{q}=\lambda (q_x-iq_y)$ is the Fourier transformed Rashba operator.
The poles of the Green's function give the spectrum of elementary excitations $\omega_{\mathbf{q}s}$.
It can be found by replacing $i\omega_n$ with $\omega_{\mathbf{q}s}$ and solving the secular equation $\det{\mathcal{G}^{-1}({\bf q},\omega_{\mathbf{q}s})}=0$. The index
$s$ denotes different solutions of the secular equation. We retain only two positive solutions in the following.
The spectrum is used to study thermodynamic properties of the system. The Gaussian action permits direct integration over the fluctuating fields to yield
the grand potential $-\ln\mathcal{Z}/\beta=S_0/\beta+\Omega_f$, where the second term

\begin{equation}
    \Omega_f=\sum_{\mathbf{q},s}
    \left[\frac{\ln{(1-e^{-\beta\omega_{\mathbf{q}s}})}}{2\beta}
            +\frac{\omega_{\mathbf{q}s}-\epsilon_\mathbf{q}}{2}+\frac{(g^2+g_{\uparrow\downarrow}^2)n_0^2}{2\mathbf{q}^2}\right]
\end{equation}
is the result of the Gaussian integration.
\begin{figure}[t]
 \includegraphics[width=0.5\textwidth]{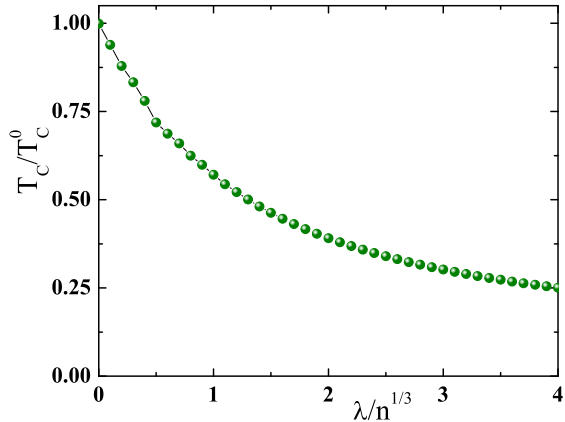}
\caption{(color online). The critical temperature of condensation as a function of the SOC strength.
SOC decreases the critical temperature due to increase of the low energy density of states on
the circle $q_\perp=\lambda/2$. Here $T_c^0$ is the critical temperature for a non-interacting Bose system without SOC. We have set $g_{\uparrow\downarrow}=g$.}
\label{Fig1}
\end{figure}
The density of particles excited out of the condensates for either species can be easily evaluated as
$n_f=1/(2L^3)\left(\partial\Omega_f/\partial \mu\right)_{\mu=\mu_p}$. At the critical temperature the condensed density $n_0=0$ and $n_f$ is equal to the density
of the total number of particles $n$. In this special case, it is straightforward to calculate analytically the poles of the Green's function and obtain the following
secular equation determining the critical temperature
\begin{equation}
n=\frac{1}{2L^3}\sum_{\mathbf{q},s=\pm}\frac{1}{{\rm e}^{[q_z^2+(q_\perp+s\lambda/2)^2]/T_c}-1}.
\end{equation}
In the absence of SOC, the two terms are identical and we get the usual number equation to determine the critical temperature of condensation. The biggest contribution
to the sum over the momentum is from the vicinity of the point $|{\bf q}|=0$. SOC changes the situation. Now the biggest contribution to the sum
comes from the circle $q_z=0, q_\perp = \lambda/2$ with much higher weight than in the previous case. The critical temperature $T_c$ must be decreased for fixed
density. More quantitatively, the critical temperature in the thermodynamic limit is estimated as
\begin{eqnarray}
   T_c\approx T_c^0\frac{g_{3/2}(1)}{g_{3/2}(z)}.
\end{eqnarray}
Here, $T_c^0=4\pi [n/g_{3/2}(1)]^{2/3}$ is the critical temperature for a non-interacting Bose gas without SOC,
$z=\exp{[-\lambda^2/(4T_c)]}$, and $g_p(z)=\sum_{l=1}^\infty z^l/l^p $\cite{LEV03}. As shown in Fig.~$\ref{Fig1}$, the critical temperature $T_c$ indeed decreases rapidly
as the SOC strength increases.
\begin{figure}[t]
 \includegraphics[width=0.5\textwidth]{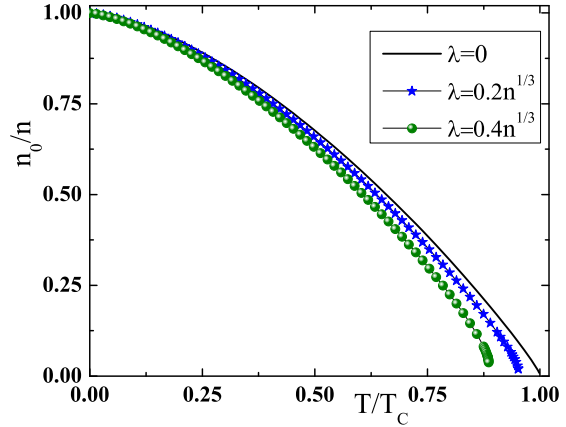}
\caption{(color online). The condensate fraction as a function of temperature for various spin-orbit coupling strengths $\lambda$. The effect of SOC
on the thermal depletion is more pronounced at higher temperatures, since the thermal component benefits more from the low energy states than the condensed one.
We have set $g_{\uparrow\downarrow}=g$.}
\label{Fig2}
\end{figure}

The situation is more complicated for temperatures smaller than the critical temperature. The secular equation should be supplemented with the equation for the total density
$n_0+n_f=n$. By solving the two equations we obtain the condensate density $n_0$ for a fixed total density of atoms $n$. Being an intrinsic property of
a Bose-Einstein condensate, the condensate fraction $n_0/n$ provides key information about the robustness of the superfluid state.
It is shown in Fig.~$\ref{Fig2}$, where three typical SOC couplings ($\lambda=0$, $\lambda/n^{1/3}=0.2$, and $\lambda/n^{1/3}=0.4$) are chosen for comparison.
In the absence of SOC, the condensate fraction decreases gradually to zero as the temperature reaches $T_c^0$,
since under the Popov approximation, the critical temperature of a weakly interacting Bose gas is identical to that of a non-interacting
one~\cite{DAF99,PET08}. The effect of SOC on the condensate fraction is more pronounced when the temperature gets closer to the transition temperature $T_c$.
We can explain this by a similar argument leading to the reduction of the critical temperature: At a fixed temperature the thermal component
benefits mainly from the low energy states living on the circle $q_\perp =\lambda/2$, while the condensed component mainly from a single momentum at $q_x=\lambda/2$.
Therefore, SOC affects the thermal component more than the condensed one, leading to the more pronounced thermal depletion at higher temperatures.


We turn now to study the effects of SOC on the system in two spatial dimensions.
In the absence of SOC there is a quasi-long range order in two dimensions at sufficiently low temperatures with correlation function decaying according to some power law.
At higher temperatures the correlation function decays exponentially, where the algebraic order is destroyed by proliferation of vortices.
The BKT separates these two regimes \cite{NAG99}. Since there is no condensation now,
the Popov approximation is not applicable. Here we derive a low energy effective theory suitable to probe the correlation function.
Anticipating that phase degrees of freedom play an essential role now, we adopt the density-phase representation of the field operators,
$\psi_\sigma=\rho_\sigma e^{i\theta_\sigma}$. The fields are then separated again into a mean-field part and a fluctuating part as
$\rho_\sigma=\rho_{0}+\delta\rho_\sigma$ and $\theta_\sigma=\theta_{0\sigma}+\delta\theta_\sigma$.
Here, $\rho_0$ is usually called a quasi-condensate density and the explicit breaking of the rotational symmetry by SOC results in $\theta_{0\uparrow}=\theta_{0\downarrow}-\pi=\lambda x/2$.
Retaining terms of zeroth and quadratic order in the fluctuating fields, the action is split into two parts $S\approx S_0+S_f$, where similarly to the previous case
$S_0=\beta L^2\left[-2(\lambda^2/4+\mu)\rho_0+(g+g_{\uparrow\downarrow})\rho_0^2\right]$ is the mean-field contribution.
The Gaussian action $S_f$ contains the fluctuating fields up to the second order. This allows us to integrate the density fluctuations $\delta\rho_\sigma$ out yieding
\begin{eqnarray}
    &&S_{f}=\int d\tau \int d\mathbf{r}\left\{\frac{(\partial_\tau\theta_+)^2}{g+g_{\uparrow\downarrow}}+\frac{(\partial_\tau\theta_- +i2\lambda\partial_x\theta_-)^2}
{g-g_{\uparrow\downarrow}+\lambda^2/(2\rho_0)}\right.\nonumber\\
    &&+\left.\frac{\rho_0}{2}\left[(\nabla\theta_+)^2+(\nabla\theta_-)^2\right]+\rho_0\lambda\theta_-\partial_y\theta_++\frac{\rho_0}{2}\lambda^2\theta_-^2\right\},
\end{eqnarray}
where $\theta_\pm =\delta\theta_\uparrow \pm \delta\theta_\downarrow$. In the absence of SOC, the above action describes two decoupled quantum XY models
representing two branches of phase fluctuations with linear spectrum in momentum space yielding the quasi-long range order.
The corresponding correlation functions
$C_\pm({\bf r})=\langle e^{[\theta_\pm({\bf r})-\theta_\pm(0)]}\rangle\propto r^{-T/\rho_0(g\pm g_{\uparrow\downarrow})}$ decay according to the power law rather than
exponentially \cite{NAG99}. In the presence of SOC, we notice that the field $\theta_-$ describes a massive fluctuation as the
action $S_g$ contains the term  $\rho_0\lambda^2\theta_-^2/2$. At low energy, we can integrate out the massive field $\theta_-$ by assuming homogeneous
temporal fluctuations to get an effective action solely in terms of the collective variable $\theta_+$ (for simplicity we set $g\approx g_{\uparrow\downarrow}$):
\begin{eqnarray}
      S_f\approx \frac{1}{2}\int d\tau \int d\mathbf{r}\left[\frac{1}{g}(\partial_\tau\theta_+)^2+\rho_0(\partial_x\theta_+)^2+\frac{\rho_0}{\lambda^2}(\partial_y^2\theta_+)^2\right].\nonumber\\
\end{eqnarray}
The first two terms in the action are the part of the familiar XY model. The third term is the manifestation of the broken rotational symmetry and it adds a twist.
To see this we calculate the spectrum of elementary excitations.
We first Fourier transform the action for it to become $S_f=1/2\sum_{{\bf q},n}(w_n^2/g+\rho_0 q_x^2+\rho_0 q_y^4/\lambda^2)|\theta_+({\bf q},w_n)|^2$.
The phase fluctuations in the Fourier space can be easily evaluated, $\langle |\theta_+(\mathbf{q},w_n)|^2 \rangle=g(w_n^2+w_{\bf q}^2)^{-1}$. The poles of this expression
gives the low energy spectrum of the symmetric phase $w_\mathbf{q}=\sqrt{g\rho_0(q_x^2+q_y^4/\lambda^2)}$. It is clear that now the energy carried by elementary
excitations of the system does not scale linearly as in the XY model case. As a result, less energy is required to excite low energy modes as compared to the XY model case,
enhancing the role of quantum fluctuations.  The Landau criterion for the superfluid critical velocity gives
${\rm min}_{\bf q}(w_{\bf q}/q) =0$, implying SOC destroys superfluidity in two dimensions. This is substantiated by the form of the correlation function,
which reads $C_+({\bf r})\approx e^{-I}$ with $I=\sum_{\mathbf{q},n}
\langle |\theta_+(\mathbf{q},w_n)|^2\rangle(1-e^{i\mathbf{q}\mathbf{r}})/2\beta$. The asymptotic behavior of it for large separations along the $x$-direction is
$I\approx T\sqrt{\lambda|x|}/(2\pi g\rho_0)$ and along the $y$-direction is $I\approx T\lambda|y|/(2g\rho_0)$. Hence, at any finite temperature,
the correlation function decays exponentially along both $x$- and $y$- directions,  in stark contrast to the
power law exhibited by the conventional XY model in the low temperature phase. The superfluid order in two dimensions is predicted to be disordered by SOC.

{\it Summary.}-- We predict dramatic implications of SOC on the thermal properties of atomic Bose gases:
In three dimensions SOC reduces the critical temperature of condensation and enhances the thermal depletion of the condensate, while
in two dimensions it destroys superfluid order at any finite temperature. This is explained by the role of quantum fluctuations amplified by SOC.
We hope that current work will stimulate experiments to verify our predictions and will add to the overall understanding of the thermal quantum fluids.

We would like to thank Zhi-Gao Huang, Xiu-Min Lin and Joachim Brand for useful discussions.
RL was supported by the NBPRC under Grant No. 2011CBA00200, the NSFC under Grants No. 11274064, and by NCET.
OF was supported by the Marsden Fund Council from Government funding (contract No. MAU1205), administrated by the Royal Society of New Zealand.


\begin{thebibliography}{10}%
\makeatletter
\providecommand \@ifxundefined [1]{%
 \ifx #1\undefined \expandafter \@firstoftwo
 \else \expandafter \@secondoftwo
\fi
}%
\providecommand \@ifnum [1]{%
 \ifnum #1\expandafter \@firstoftwo
 \else \expandafter \@secondoftwo
\fi
}%
\providecommand \enquote [1]{``#1''}%
\providecommand \bibnamefont  [1]{#1}%
\providecommand \bibfnamefont [1]{#1}%
\providecommand \citenamefont [1]{#1}%
\providecommand\href[0]{\@sanitize\@href}%
\providecommand\@href[1]{\endgroup\@@startlink{#1}\endgroup\@@href}%
\providecommand\@@href[1]{#1\@@endlink}%
\providecommand \@sanitize [0]{\begingroup\catcode`\&12\catcode`\#12\relax}%
\@ifxundefined \pdfoutput {\@firstoftwo}{%
 \@ifnum{\z@=\pdfoutput}{\@firstoftwo}{\@secondoftwo}%
}{%
 \providecommand\@@startlink[1]{\leavevmode}%
 \providecommand\@@endlink[0]{}%
}{%
 \providecommand\@@startlink[1]{%
  \leavevmode
  \pdfstartlink
   attr{/Border[0 0 1 ]/H/I/C[0 1 1]}%
   user{/Subtype/Link/A<</Type/Action/S/URI/URI(#1)>>}%
  \relax
 }%
 \providecommand\@@endlink[0]{\pdfendlink}%
}%
\providecommand \url  [0]{\begingroup\@sanitize \@url }%
\providecommand \@url [1]{\endgroup\@href {#1}{\urlprefix}}%
\providecommand \urlprefix [0]{URL }%
\providecommand \Eprint[0]{\href }%
\@ifxundefined \urlstyle {%
  \providecommand \doi [1]{doi:\discretionary{}{}{}#1}%
}{%
  \providecommand \doi [0]{doi:\discretionary{}{}{}\begingroup
  \urlstyle{rm}\Url }%
}%
\providecommand \doibase [0]{http://dx.doi.org/}%
\providecommand \Doi[1]{\href{\doibase#1}}%
\providecommand \bibAnnote [3]{%
  \BibitemShut{#1}%
  \begin{quotation}\noindent
    \textsc{Key:}\ #2\\\textsc{Annotation:}\ #3%
  \end{quotation}%
}%
\providecommand \bibAnnoteFile [2]{%
  \IfFileExists{#2}{\bibAnnote {#1} {#2} {\input{#2}}}{}%
}%
\providecommand \typeout [0]{\immediate \write \m@ne }%
\providecommand \selectlanguage [0]{\@gobble}%
\providecommand \bibinfo [0]{\@secondoftwo}%
\providecommand \bibfield [0]{\@secondoftwo}%
\providecommand \translation [1]{[#1]}%
\providecommand \BibitemOpen[0]{}%
\providecommand \bibitemStop [0]{}%
\providecommand \bibitemNoStop [0]{.\EOS\space}%
\providecommand \EOS [0]{\spacefactor3000\relax}%
\providecommand \BibitemShut [1]{\csname bibitem#1\endcsname}%
\bibitem{XIA}%
  \BibitemOpen
  \bibfield{author}{%
  \bibinfo {author} {\bibfnamefont{D.}~\bibnamefont{Xiao}}, \bibinfo {author}
  {\bibfnamefont{M.-C.}\ \bibnamefont{Chang}},\ and\ \bibinfo {author}
  {\bibfnamefont{Q.}~\bibnamefont{Niu}},\ }%
  \bibfield{journal}{%
  \bibinfo {journal} {Rev. Mod. Phys.}\ }%
  \textbf{\bibinfo {volume} {82}},\ \bibinfo {note} {1959 (2010), M. Z. Hasan
  and C. L. Kane, ibid. \textbf{82}, 3045 (2010); X. L. Qi and S. C. Zhang,
  ibid. \textbf{83}, 1057 (2011).}%
  \bibAnnoteFile{Stop}{XIA}%
\bibitem{MOU12}%
  \BibitemOpen
  \bibfield{author}{%
  \bibinfo {author} {\bibfnamefont{V.}~\bibnamefont{Mourik}}, \bibinfo {author}
  {\bibfnamefont{K.}~\bibnamefont{Zuo}}, \bibinfo {author}
  {\bibfnamefont{S.~M.}\ \bibnamefont{Frlov}}, \bibinfo {author}
  {\bibfnamefont{S.~R.}\ \bibnamefont{Plissard}}, \bibinfo {author}
  {\bibnamefont{E.P.A.M.Bakkers}},\ and\ \bibinfo {author}
  {\bibfnamefont{L.~P.}\ \bibnamefont{Kouwenhoven}},\ }%
  \bibfield{journal}{%
  \bibinfo {journal} {Science}\ }%
  \textbf{\bibinfo {volume} {336}},\ \bibinfo {pages} {1003} (\bibinfo {year}
  {2012})%
  \bibAnnoteFile{NoStop}{MOU12}%
\bibitem{ZOL05}%
  \BibitemOpen
  \bibfield{author}{%
  \bibinfo {author} {\bibfnamefont{D.}~\bibnamefont{Jaksch}}\ and\ \bibinfo
  {author} {\bibfnamefont{P.}~\bibnamefont{Zoller}},\ }%
  \bibfield{journal}{%
  \bibinfo {journal} {Annals of Physics}\ }%
  \textbf{\bibinfo {volume} {315}},\ \bibinfo {pages} {52} (\bibinfo {year}
  {2005})%
  \bibAnnoteFile{NoStop}{ZOL05}%
\bibitem{LIN}%
  \BibitemOpen
  \bibfield{author}{%
  \bibinfo {author} {\bibfnamefont{Y.-J.}\ \bibnamefont{Lin}}, \bibinfo
  {author} {\bibfnamefont{R.~L.}\ \bibnamefont{Compton}}, \bibinfo {author}
  {\bibfnamefont{A.~R.}\ \bibnamefont{Perry}}, \bibinfo {author}
  {\bibfnamefont{W.~D.}\ \bibnamefont{Phillips}}, \bibinfo {author}
  {\bibfnamefont{J.~V.}\ \bibnamefont{Porto}},\ and\ \bibinfo {author}
  {\bibfnamefont{I.~B.}\ \bibnamefont{Spielman}},\ }%
  \bibfield{journal}{%
  \bibinfo {journal} {Phys. Rev. Lett.}\ }%
  \textbf{\bibinfo {volume} {102}},\ \bibinfo {note} {130401 (2009); Y.-J. Lin
  and R. L. Compton and K. Jim\'{e}nez-Garcia and J. V. Porto and I. B.
  Spielman, Nature \textbf{462}, 628 (2009); Y.-J. Lin and K.
  Jim\'{e}nez-Garcia and I. B. Spielman, Nature \textbf{471}, 83 (2011).}%
  \bibAnnoteFile{Stop}{LIN}%
\bibitem{BLO11}%
  \BibitemOpen
  \bibfield{author}{%
  \bibinfo {author} {\bibfnamefont{M.}~\bibnamefont{Aidelsburger}}, \bibinfo
  {author} {\bibfnamefont{M.}~\bibnamefont{Atala}}, \bibinfo {author}
  {\bibfnamefont{S.}~\bibnamefont{Nascimb\'{e}ne}}, \bibinfo {author}
  {\bibfnamefont{S.}~\bibnamefont{Trotzky}}, \bibinfo {author}
  {\bibfnamefont{Y.-A.}\ \bibnamefont{Chen}},\ and\ \bibinfo {author}
  {\bibfnamefont{I.}~\bibnamefont{Bloch}},\ }%
  \bibfield{journal}{%
  \bibinfo {journal} {Phys. Rev. Lett.}\ }%
  \textbf{\bibinfo {volume} {107}},\ \bibinfo {pages} {255301} (\bibinfo {year}
  {2011})%
  \bibAnnoteFile{NoStop}{BLO11}%
\bibitem{CHE12}%
  \BibitemOpen
  \bibfield{author}{%
  \bibinfo {author} {\bibfnamefont{J.}~\bibnamefont{Zhang}}, \bibinfo {author}
  {\bibfnamefont{S.}~\bibnamefont{Ji}}, \bibinfo {author}
  {\bibfnamefont{Z.}~\bibnamefont{Chen}}, \bibinfo {author}
  {\bibfnamefont{L.}~\bibnamefont{Zhang}}, \bibinfo {author}
  {\bibfnamefont{Z.}~\bibnamefont{Du}}, \bibinfo {author}
  {\bibfnamefont{B.}~\bibnamefont{Yan}}, \bibinfo {author}
  {\bibfnamefont{G.}~\bibnamefont{Pan}}, \bibinfo {author}
  {\bibfnamefont{B.}~\bibnamefont{Zhao}}, \bibinfo {author}
  {\bibfnamefont{Y.}~\bibnamefont{Deng}}, \bibinfo {author}
  {\bibfnamefont{H.}~\bibnamefont{Zhai}}, \bibinfo {author}
  {\bibfnamefont{S.}~\bibnamefont{Chen}},\ and\ \bibinfo {author}
  {\bibfnamefont{J.}~\bibnamefont{Pan}},\ }%
  \bibinfo {journal} {arxiv: 1201.6018}%
  \bibAnnoteFile{NoStop}{CHE12}%
\bibitem{ZHA12}%
  \BibitemOpen
\bibfield{journal}{%
    }%
  \bibfield{author}{%
  \bibinfo {author} {\bibfnamefont{P.}~\bibnamefont{Wang}}, \bibinfo {author}
  {\bibfnamefont{Z.-Q.}\ \bibnamefont{Yu}}, \bibinfo {author}
  {\bibfnamefont{Z.}~\bibnamefont{Fu}}, \bibinfo {author}
  {\bibfnamefont{J.}~\bibnamefont{Miao}}, \bibinfo {author}
  {\bibfnamefont{L.}~\bibnamefont{Huang}}, \bibinfo {author}
  {\bibfnamefont{S.}~\bibnamefont{Chai}}, \bibinfo {author}
  {\bibfnamefont{H.}~\bibnamefont{Zhai}},\ and\ \bibinfo {author}
  {\bibfnamefont{J.}~\bibnamefont{Zhang}},\ }%
  \bibfield{journal}{%
  \bibinfo {journal} {Phys. Rev. Lett.}\ }%
  \textbf{\bibinfo {volume} {109}},\ \bibinfo {pages} {095301} (\bibinfo {year}
  {2012})%
  \bibAnnoteFile{NoStop}{ZHA12}%
\bibitem{ZWI12}%
  \BibitemOpen
  \bibfield{author}{%
  \bibinfo {author} {\bibfnamefont{L.~W.}\ \bibnamefont{Cheuk}}, \bibinfo
  {author} {\bibfnamefont{A.~T.}\ \bibnamefont{Sommer}}, \bibinfo {author}
  {\bibfnamefont{Z.}~\bibnamefont{Hadzibabic}}, \bibinfo {author}
  {\bibfnamefont{T.}~\bibnamefont{Yefsah}}, \bibinfo {author}
  {\bibfnamefont{W.~S.}\ \bibnamefont{Bakr}},\ and\ \bibinfo {author}
  {\bibfnamefont{M.~W.}\ \bibnamefont{Zwierlein}},\ }%
  \bibfield{journal}{%
  \bibinfo {journal} {Phys. Rev. Lett.}\ }%
  \textbf{\bibinfo {volume} {109}},\ \bibinfo {pages} {095302} (\bibinfo {year}
  {2012})%
  \bibAnnoteFile{NoStop}{ZWI12}%
\bibitem{STA08}%
  \BibitemOpen
  \bibfield{author}{%
  \bibinfo {author} {\bibfnamefont{T.~D.}\ \bibnamefont{Stanescu}}, \bibinfo
  {author} {\bibfnamefont{B.}~\bibnamefont{Anderson}},\ and\ \bibinfo {author}
  {\bibfnamefont{V.}~\bibnamefont{Galitski}},\ }%
  \bibfield{journal}{%
  \bibinfo {journal} {Phys. Rev. A}\ }%
  \textbf{\bibinfo {volume} {78}},\ \bibinfo {pages} {023616} (\bibinfo {year}
  {2008})%
  \bibAnnoteFile{NoStop}{STA08}%
\bibitem{ZHA10}%
  \BibitemOpen
  \bibfield{author}{%
  \bibinfo {author} {\bibfnamefont{C.}~\bibnamefont{Wang}}, \bibinfo {author}
  {\bibfnamefont{C.}~\bibnamefont{Gao}}, \bibinfo {author}
  {\bibfnamefont{C.~M.}\ \bibnamefont{Jian}},\ and\ \bibinfo {author}
  {\bibfnamefont{H.}~\bibnamefont{Zhai}},\ }%
  \bibfield{journal}{%
  \bibinfo {journal} {Phys. Rev. Lett.}\ }%
  \textbf{\bibinfo {volume} {105}},\ \bibinfo {pages} {160403} (\bibinfo {year}
  {2010})%
  \bibAnnoteFile{NoStop}{ZHA10}%
\bibitem{CHA11}%
  \BibitemOpen
  \bibfield{author}{%
  \bibinfo {author} {\bibfnamefont{C.-M.}\ \bibnamefont{Jian}}\ and\ \bibinfo
  {author} {\bibfnamefont{H.}~\bibnamefont{Zhai}},\ }%
  \bibfield{journal}{%
  \bibinfo {journal} {Phys. Rev. B}\ }%
  \textbf{\bibinfo {volume} {84}},\ \bibinfo {pages} {060508} (\bibinfo {year}
  {2011})%
  \bibAnnoteFile{NoStop}{CHA11}%
\bibitem{TIN11}%
  \BibitemOpen
  \bibfield{author}{%
  \bibinfo {author} {\bibfnamefont{T.-L.}\ \bibnamefont{Ho}}\ and\ \bibinfo
  {author} {\bibfnamefont{S.}~\bibnamefont{Zhang}},\ }%
  \bibfield{journal}{%
  \bibinfo {journal} {Phys. Rev. Lett.}\ }%
  \textbf{\bibinfo {volume} {107}},\ \bibinfo {pages} {150403} (\bibinfo {year}
  {2011})%
  \bibAnnoteFile{NoStop}{TIN11}%
\bibitem{YUN}%
  \BibitemOpen
  \bibfield{author}{%
  \bibinfo {author} {\bibfnamefont{Y.}~\bibnamefont{Li}}, \bibinfo {author}
  {\bibfnamefont{L.~P.}\ \bibnamefont{Pitaevskii}},\ and\ \bibinfo {author}
  {\bibfnamefont{S.}~\bibnamefont{Stringari}},\ }%
  \bibfield{journal}{%
  \bibinfo {journal} {Phys. Rev. Lett.}\ }%
  \textbf{\bibinfo {volume} {108}},\ \bibinfo {note} {225301 (2012); Y. Li, G.
  I. Martone, L. P. Pitaevskii, and S. Stringari, ibid. \textbf{110}, 235302
  (2013).}%
  \bibAnnoteFile{Stop}{YUN}%
\bibitem{AND}%
  \BibitemOpen
  \bibfield{author}{%
  \bibinfo {author} {\bibfnamefont{B.~M.}\ \bibnamefont{Anderson}}, \bibinfo
  {author} {\bibfnamefont{G.}~\bibnamefont{Juzeli\={u}nas}}, \bibinfo {author}
  {\bibfnamefont{V.~M.}\ \bibnamefont{Galitski}},\ and\ \bibinfo {author}
  {\bibfnamefont{I.~B.}\ \bibnamefont{Spielman}},\ }%
  \bibfield{journal}{%
  \bibinfo {journal} {Phys. Rev. Lett.}\ }%
  \textbf{\bibinfo {volume} {108}},\ \bibinfo {note} {235301 (2012); B. M.
  Anderson, I. B. Spielman and G. Juzeli\={u}nas, Phys. Rev. Lett.
  \textbf{111}, 125301 (2013).}%
  \bibAnnoteFile{Stop}{AND}%
\bibitem{FER}%
  \BibitemOpen
  \bibfield{author}{%
  \bibinfo {author} {\bibfnamefont{M.}~\bibnamefont{Iskin}}\ and\ \bibinfo
  {author} {\bibfnamefont{A.~L.}\ \bibnamefont{Subasi}},\ }%
  \bibfield{journal}{%
  \bibinfo {journal} {Phys. Rev. Lett.}\ }%
  \textbf{\bibinfo {volume} {107}},\ \bibinfo {note} {050402 (2011), M. Gong,
  S. Tewari, and C. Zhang, ibid. \textbf{107}, 195303 (2011); H. Hu, L. Jiang,
  X.-J. Liu, and H. Pu, ibid. \textbf{107}, 195304 (2011); Z.-Q. Yu and H.
  Zhai, ibid. \textbf{107},195305 (2011); R. Liao, Y. Yi-Xiang and W-M. Liu,
  ibid.\textbf{108}, 080406(2012).}%
  \bibAnnoteFile{Stop}{FER}%
\bibitem{BOSE}%
  \BibitemOpen
  \bibfield{author}{%
  \bibinfo {author} {\bibfnamefont{S.}~\bibnamefont{Sinha}}, \bibinfo {author}
  {\bibfnamefont{R.}~\bibnamefont{Nath}},\ and\ \bibinfo {author}
  {\bibfnamefont{L.}~\bibnamefont{Santos}},\ }%
  \bibfield{journal}{%
  \bibinfo {journal} {Phys. Rev. Lett.}\ }%
  \textbf{\bibinfo {volume} {107}},\ \bibinfo {note} {270401 (2011); H. Hu, B.
  Ramachandhran, H. Pu, and X.-J. Liu, Phys. Rev. Lett. \textbf{108},010402
  (2012); T. Kawakami, T. Mizushima, M. Nitta, and K. Machida, Phys. Rev. Lett.
  \textbf{109},015301 (2012); Z. F. Xu, Y. Kawaguchi, L. You, and M. Ueda,
  Phys. Rev. A 86, 033628 (2012).}%
  \bibAnnoteFile{Stop}{BOSE}%
\bibitem{LIU12}%
  \BibitemOpen
  \bibfield{author}{%
  \bibinfo {author} {\bibfnamefont{H.}~\bibnamefont{Hu}}\ and\ \bibinfo
  {author} {\bibfnamefont{X.-J.}\ \bibnamefont{Liu}},\ }%
  \bibfield{journal}{%
  \bibinfo {journal} {Phys. Rev. A}\ }%
  \textbf{\bibinfo {volume} {85}},\ \bibinfo {pages} {013619} (\bibinfo {year}
  {2012})%
  \bibAnnoteFile{NoStop}{LIU12}%
\bibitem{QI13}%
  \BibitemOpen
  \bibfield{author}{%
  \bibinfo {author} {\bibfnamefont{Q.}~\bibnamefont{Zhou}}\ and\ \bibinfo
  {author} {\bibfnamefont{X.}~\bibnamefont{Cui}},\ }%
  \bibfield{journal}{%
  \bibinfo {journal} {Phys. Rev. Lett.}\ }%
  \textbf{\bibinfo {volume} {110}},\ \bibinfo {pages} {140407} (\bibinfo {year}
  {2013})%
  \bibAnnoteFile{NoStop}{QI13}%
\bibitem{BAR12}%
  \BibitemOpen
  \bibfield{author}{%
  \bibinfo {author} {\bibfnamefont{R.}~\bibnamefont{Barnett}}, \bibinfo
  {author} {\bibfnamefont{S.}~\bibnamefont{Powell}}, \bibinfo {author}
  {\bibfnamefont{T.}~\bibnamefont{Grass}}, \bibinfo {author}
  {\bibfnamefont{M.}~\bibnamefont{Lewenstein}},\ and\ \bibinfo {author}
  {\bibfnamefont{S.}~\bibnamefont{DasSarma}},\ }%
  \bibfield{journal}{%
  \bibinfo {journal} {Phys. Rev. A}\ }%
  \textbf{\bibinfo {volume} {85}},\ \bibinfo {pages} {023615} (\bibinfo {year}
  {2012})%
  \bibAnnoteFile{NoStop}{BAR12}%
\bibitem{CUI12}%
  \BibitemOpen
  \bibfield{author}{%
  \bibinfo {author} {\bibfnamefont{X.}~\bibnamefont{Cui}}\ and\ \bibinfo
  {author} {\bibfnamefont{Q.}~\bibnamefont{Zhou}},\ }%
  \bibfield{journal}{%
  \bibinfo {journal} {Phys. Rev. A}\ }%
  \textbf{\bibinfo {volume} {87}},\ \bibinfo {pages} {031604} (\bibinfo {year}
  {2013})%
  \bibAnnoteFile{NoStop}{CUI12}%
\bibitem{BAY12}%
  \BibitemOpen
  \bibfield{author}{%
  \bibinfo {author} {\bibfnamefont{T.}~\bibnamefont{Ozawa}}\ and\ \bibinfo
  {author} {\bibfnamefont{G.}~\bibnamefont{Baym}},\ }%
  \bibfield{journal}{%
  \bibinfo {journal} {Phys. Rev. Lett.}\ }%
  \textbf{\bibinfo {volume} {109}},\ \bibinfo {pages} {025301} (\bibinfo {year}
  {2012})%
  \bibAnnoteFile{NoStop}{BAY12}%
\bibitem{PAN13}%
  \BibitemOpen
  \bibfield{author}{%
  \bibinfo {author} {\bibfnamefont{J.-Y.}\ \bibnamefont{Zhang}}, \bibinfo
  {author} {\bibfnamefont{S.-C.}\ \bibnamefont{Ji}}, \bibinfo {author}
  {\bibfnamefont{L.}~\bibnamefont{Zhang}}, \bibinfo {author}
  {\bibfnamefont{Z.-D.}\ \bibnamefont{Du}}, \bibinfo {author}
  {\bibfnamefont{W.}~\bibnamefont{Zheng}}, \bibinfo {author}
  {\bibfnamefont{Y.-J.}\ \bibnamefont{Deng}}, \bibinfo {author}
  {\bibfnamefont{H.}~\bibnamefont{Zhai}}, \bibinfo {author}
  {\bibfnamefont{S.}~\bibnamefont{Chen}},\ and\ \bibinfo {author}
  {\bibfnamefont{J.-W.}\ \bibnamefont{Pan}}\ }%
  \bibinfo {note} {arxiv:1305.7054}%
  \bibAnnoteFile{NoStop}{PAN13}%
\bibitem{SIM06}%
  \BibitemOpen
  \bibfield{author}{%
  \bibinfo {author} {\bibfnamefont{A.}~\bibnamefont{Altland}}\ and\ \bibinfo
  {author} {\bibfnamefont{B.}~\bibnamefont{Simons}},\ }%
  \emph{\bibinfo {title} {Condensed Matter Field Theory}}\ (\bibinfo
  {publisher} {Cambridge University Press},\ \bibinfo {address} {Cambridge},\
  \bibinfo {year} {2006})%
  \bibAnnoteFile{NoStop}{SIM06}%
\bibitem{GRI96}%
  \BibitemOpen
  \bibfield{author}{%
  \bibinfo {author} {\bibfnamefont{A.}~\bibnamefont{Griffin}},\ }%
  \bibfield{journal}{%
  \bibinfo {journal} {Phys. Rev. B}\ }%
  \textbf{\bibinfo {volume} {53}},\ \bibinfo {pages} {9341} (\bibinfo {year}
  {1996})%
  \bibAnnoteFile{NoStop}{GRI96}%
\bibitem{HUG59}%
  \BibitemOpen
  \bibfield{author}{%
  \bibinfo {author} {\bibfnamefont{N.}~\bibnamefont{Hugenholtz}}\ and\ \bibinfo
  {author} {\bibfnamefont{D.}~\bibnamefont{Pines}},\ }%
  \bibfield{journal}{%
  \bibinfo {journal} {Phys. Rev.}\ }%
  \textbf{\bibinfo {volume} {116}},\ \bibinfo {pages} {489} (\bibinfo {year}
  {1959})%
  \bibAnnoteFile{NoStop}{HUG59}%
\bibitem{GOL62}%
  \BibitemOpen
  \bibfield{author}{%
  \bibinfo {author} {\bibfnamefont{J.}~\bibnamefont{Goldstone}},\ }%
  \bibfield{journal}{%
  \bibinfo {journal} {Phys. Rev.}\ }%
  \textbf{\bibinfo {volume} {127}},\ \bibinfo {pages} {2} (\bibinfo {year}
  {1962})%
  \bibAnnoteFile{NoStop}{GOL62}%
\bibitem{POP01}%
  \BibitemOpen
  \bibfield{author}{%
  \bibinfo {author} {\bibfnamefont{V.}~\bibnamefont{Popov}},\ }%
  \emph{\bibinfo {title} {Functional integrals in quantum field theory and
  stastical physics}}\ (\bibinfo {publisher} {Reidel},\ \bibinfo {address}
  {Dordrecht},\ \bibinfo {year} {2001})%
  \bibAnnoteFile{NoStop}{POP01}%
\bibitem{AND04}%
  \BibitemOpen
  \bibfield{author}{%
  \bibinfo {author} {\bibfnamefont{J.~O.}\ \bibnamefont{Anderson}},\ }%
  \bibfield{journal}{%
  \bibinfo {journal} {Rev. Mod. Phys.}\ }%
  \textbf{\bibinfo {volume} {76}},\ \bibinfo {pages} {599} (\bibinfo {year}
  {2004})%
  \bibAnnoteFile{NoStop}{AND04}%
\bibitem{LEV03}%
  \BibitemOpen
  \bibfield{author}{%
  \bibinfo {author} {\bibfnamefont{L.}~\bibnamefont{Pitaevskii}}\ and\ \bibinfo
  {author} {\bibfnamefont{S.}~\bibnamefont{Stringari}},\ }%
  \emph{\bibinfo {title} {Bose-Einstein Condensation}}\ (\bibinfo {publisher}
  {Oxford},\ \bibinfo {address} {New York},\ \bibinfo {year} {2003})\ \bibinfo
  {note} {page 17}%
  \bibAnnoteFile{NoStop}{LEV03}%
\bibitem{DAF99}%
  \BibitemOpen
  \bibfield{author}{%
  \bibinfo {author} {\bibfnamefont{F.}~\bibnamefont{Dalfovo}}, \bibinfo
  {author} {\bibfnamefont{S.}~\bibnamefont{Giorgini}}, \bibinfo {author}
  {\bibfnamefont{L.~P.}\ \bibnamefont{Pitaevskii}},\ and\ \bibinfo {author}
  {\bibfnamefont{S.}~\bibnamefont{Stringari}},\ }%
  \bibfield{journal}{%
  \bibinfo {journal} {Rev. Mod. Phys.}\ }%
  \textbf{\bibinfo {volume} {71}},\ \bibinfo {pages} {463} (\bibinfo {year}
  {1999})%
  \bibAnnoteFile{NoStop}{DAF99}%
\bibitem{PET08}%
  \BibitemOpen
  \bibfield{author}{%
  \bibinfo {author} {\bibfnamefont{C.}~\bibnamefont{Pethick}}\ and\ \bibinfo
  {author} {\bibfnamefont{H.}~\bibnamefont{Smith}},\ }%
  \emph{\bibinfo {title} {Bose-Einstein Condensation in Dilute Gases}}\
  (\bibinfo {publisher} {Cambridge University Press},\ \bibinfo {address}
  {Cambridge},\ \bibinfo {year} {2008})%
  \bibAnnoteFile{NoStop}{PET08}%
\bibitem{NAG99}%
  \BibitemOpen
  \bibfield{author}{%
  \bibinfo {author} {\bibfnamefont{N.}~\bibnamefont{Nagaosa}},\ }%
  \emph{\bibinfo {title} {Quantum Field Theory in Condensed Matter Physics}}\
  (\bibinfo {publisher} {Springer},\ \bibinfo {address} {Berlin},\ \bibinfo
  {year} {1999})%
  \bibAnnoteFile{NoStop}{NAG99}%
\end{thebibliography}

\bibliographystyle{apsrev4-1}

\end{document}